\documentstyle[preprint,aps]{revtex}

\def\ni{\noindent}

\def\nl{\newline}

\def\ket{\rangle}
\def\ox{\otimes}

\def\-{\leftarrow}

\def\01{\{0,1\}}
\def\I{\makebox[3.75mm]{$I$}}
\def\Y{\makebox[3.75mm]{$Y$}}
\def\Z{\makebox[3.75mm]{$Z$}}

\begin{document}

\draft

\preprint{CALT-68-2067, QUIC-96-002}

\title{Efficient Computations of Encodings \\ for Quantum Error Correction}

\author{Richard Cleve\thanks{{\tt cleve@cpsc.ucalgary.ca}}}
\address{Department of Computer Science, University of Calgary,\\ 
Calgary, Alberta, Canada T2N 1N4}
\author{Daniel Gottesman\thanks{{\tt gottesma@theory.caltech.edu}}}
\address{California Institute of Technology, Pasadena, CA 91125}


\maketitle

\begin{abstract}
We show how, given any set of generators of the stabilizer of a 
quantum code, an efficient gate array that computes the codewords 
can be constructed.
For an $n$-qubit code whose stabilizer has $d$ generators, the resulting 
gate array consists of $O(n d)$ operations, and converts $k$-qubit data 
(where $k = n-d$) into $n$-qubit codewords.
\end{abstract}


\section{Introduction}

Recently, significant progress has been made in the development of 
error-correction schemes for quantum information systems 
\cite{Shor,CaSh,St1,LaMiPaZu,VaGoWi,ShSm,BeDiSmWo,Gottesman,Ca,St2}.
This includes methods for converting classical error-correcting codes 
to into quantum error-correcting codes \cite{CaSh,St1}, formalizations 
of necessary and sufficient conditions for sets of states to form 
quantum codes \cite{EkMa,BeDiSmWo,KnLa}, and a mathematical framework for a 
large class of quantum codes, known as {\em stabilizer codes} 
\cite{Gottesman,Ca}.

In order to actually use quantum codes in quantum information systems, 
constructive methods for performing encodings, error-correction, and 
decodings are required.
Towards this end, gate arrays that perform these computations are helpful.
Methods for producing gate arrays that perform error-correction for any 
stabilizer code have been presented \cite{DiSh}.
For computing encodings, the only gate arrays that have been proposed 
apply either to one specific code (such as one that encodes one qubit 
as five qubits and protects against a one-qubit error) 
\cite{LaMiPaZu,BeDiSmWo,Braunstein}, or to restricted classes of stabilizer 
codes \cite{St1,St2}.
In the present paper, we show how to efficiently construct a gate array 
that computes encodings for {\it any} stabilizer code.
In the case of an $n$-qubit code defined in terms of $d$ generators, 
our gate array consists of at most $n d$ operations (which are all one- or 
two-qubit gates and performed in-place), and it converts $k$-qubit data 
(where $k = n-d$) into $n$-qubit codewords.
The gate array can be used for decoding by running it backwards (on a 
correct codeword).
The method is illustrated throughout the paper using an $n=8$, $d=5$, $k=3$ 
code (that can handle a one-qubit error) \cite{Gottesman,Ca,St2}; 
however, it works for any stabilizer code.  

We conjecture that the sizes of our gate arrays are asymptotically optimal 
for {\it general\/} stabilizer codes, though they may not be optimal for 
some specific stabilizer codes.

In Section II, we review the basics of stabilizer codes, and introduce 
new terminology that will be useful later.
In Section III, we explain how to use Gaussian elimination to convert 
codes into a ``standard form.''
In Section IV, we show how to produce a gate array from a code in standard 
form.

\section{Stabilizer Codes}

In \cite{Gottesman,Ca} it has been shown that many quantum codes 
can be described in terms of their {\em stabilizers}.
The stabilizer of a code is the set of all operators that: \nl
(a) are formed by taking tensor products of matrices of the form 
$$I = \pmatrix{ 1 & \ 0 \cr 0 & \ 1 }, 
\ \ 
X = \pmatrix{ 0 & \ 1 \cr 1 & \ 0 }, 
\ \ 
Z = \pmatrix{ 1 & \ 0 \cr 0 & -1 }, 
\ \ 
Y = X \cdot 
Z = \pmatrix{ 0 & -1 \cr 1 & \ 0 }; \ \ \mbox{and}$$
(b) fix every codeword. \nl
Thus, the codewords form the 
$+1$-eigenspace of the operators in the stabilizer.
An error that anticommutes
with an operator $M$ in the stabilizer will take a codeword from the 
$+1$-eigenspace of $M$ to the $-1$-eigenspace of $M$.  The new state can be
distinguished from the old state, and hopefully corrected, by measuring the
eigenvalue of $M$.  If the operators in the stabilizer are chosen carefully,
a large class of errors can be corrected --- for instance, all errors 
operating on a single qubit or up to $t$ qubits.

The stabilizer of a code is always a group.  It can be most
easily described by a set of {\em generators\/} $G_1,\ldots,G_d$.
The other elements of the stabilizer are products of various $G_i$'s.
The actual choice of generators is somewhat arbitrary, as long as none of 
them is the product of other generators.  Any tensor product of 
$I$, $X$, $Y$, and $Z$ will square to $\pm 1$, but in order to have 
eigenvalues $+1$, the generators must actually square to $+1$.
In addition, in order to have a nontrivial joint $+1$-eigenspace, 
the generators must commute with each other, so the stabilizer is 
an Abelian group.

Generators for an eight qubit code that protects a three qubit state 
with up to one error 
(as explained in \cite{Gottesman,Ca,St2}) are 
\begin{eqnarray}
G_1 & = & X \ox X \ox X \ox X \ox X \ox X \ox X \ox X \nonumber \\
G_2 & = & \Z \ox \Z \ox \Z \ox \Z \ox \Z \ox \Z \ox \Z \ox \Z \nonumber \\
G_3 & = & X \ox \I \ox X \ox \I \ox \Z \ox \Y \ox \Z \ox \Y \label{eight0}\\
G_4 & = & X \ox \I \ox \Y \ox \Z \ox X \ox \I \ox \Y \ox \Z \nonumber \\
G_5 & = & X \ox \Z \ox \I \ox \Y \ox \I \ox \Y \ox X \ox \Z. \nonumber
\end{eqnarray}

Once we know the generators $G_1, \ldots, G_d$, we have most of the vital 
information about the code.
If there are $n$ qubits and $d$ generators, we can encode $k = n-d$ 
data qubits.
By measuring the eigenvalue of each of the $d$ generators, we can learn 
what, if any, error has occured, and fix it.
In \cite{DiSh}, it is explicitly shown how to do this in a 
fault-tolerant way.
If we do not require strict fault-tolerance, the array to identify the 
error syndrome has $O(n d)$ gates.
Once the error syndrome is determined, it may take a long classical 
computation to determine the actual error, depending on the code used.

In order to actually encode states using a quantum code, we need to 
decide which states will act as basis states for the coding space.
In order to do this, it is convenient to use the language of binary 
vector spaces, as in \cite{Ca}.
Define the {\it $X$-vector} of the generator 
$$G_i = G_{i1} \otimes G_{i2} \otimes \cdots \otimes G_{in},$$
as the $n$-bit vector, denoted as $X_{G_i}$, where 
$$\left(X_{G_i}\right)_{j} = 
\cases{1 & if $G_{ij} =$ $X$ or $Y$ \vspace*{-2mm}\cr
       0 & if $G_{ij} =$ $I$ or $Z$. \cr}$$
The {\it $Z$-vector} of $G_i$, denoted as $Z_{G_i}$, is defined as 
$$\left(X_{G_i}\right)_{j} = 
\cases{1 & if $G_{ij} =$ $Z$ or $Y$ \vspace*{-2mm}\cr
       0 & if $G_{ij} =$ $I$ or $X$. \cr}$$
Also, the {\it $X$-matrix} of generators $G_1, \ldots, G_d$ is defined 
as the $n \times d$ matrix, denoted as $X_G$, where 
$$\left(X_G\right)_{ji} = 
\cases{1 & if $G_{ij} =$ $X$ or $Y$ \vspace*{-2mm}\cr
       0 & if $G_{ij} =$ $I$ or $Z$, \cr}$$ 
(that is, the matrix whose columns are $X_{G_1},\ldots,X_{G_d}$).
The {\it $Z$-matrix} of $G_1, \ldots, G_d$, denoted as $Z_G$, 
is defined similarly.
Note that the $X$- and $Z$-matrices together completely determine the 
sequence of generators that they correspond to.

The $X$- and $Z$-matrices for the aforementioned generators (\ref{eight0}) 
of the eight-qubit code are 
\begin{equation}
X_G = 
\pmatrix{1 & 0 & 1 & 1 & 1 \vspace*{-4mm} \cr
         1 & 0 & 0 & 0 & 0 \vspace*{-4mm} \cr
         1 & 0 & 1 & 1 & 0 \vspace*{-4mm} \cr
         1 & 0 & 0 & 0 & 1 \vspace*{-4mm} \cr
         1 & 0 & 0 & 1 & 0 \vspace*{-4mm} \cr
         1 & 0 & 1 & 0 & 1 \vspace*{-4mm} \cr
         1 & 0 & 0 & 1 & 1 \vspace*{-4mm} \cr
         1 & 0 & 1 & 0 & 0 \cr}
\hspace*{25mm}
Z_G = 
\pmatrix{0 & 1 & 0 & 0 & 0 \vspace*{-4mm} \cr
         0 & 1 & 0 & 0 & 1 \vspace*{-4mm} \cr
         0 & 1 & 0 & 1 & 0 \vspace*{-4mm} \cr
         0 & 1 & 0 & 1 & 1 \vspace*{-4mm} \cr
         0 & 1 & 1 & 0 & 0 \vspace*{-4mm} \cr
         0 & 1 & 1 & 0 & 1 \vspace*{-4mm} \cr
         0 & 1 & 1 & 1 & 0 \vspace*{-4mm} \cr
         0 & 1 & 1 & 1 & 1 \cr}.
\label{eight1}
\end{equation}
Note that columns 1, 3, 4, 5 of $X_G$ are linearly independent, while 
column 2 is null.
In general, we will call generators whose $X$-vectors are linearly 
independent 
{\em primary} generators (``type 1'' in the language of \cite{Gottesman}), 
and generators whose $X$-vectors are null {\em secondary} generators 
(``type 2'' in the language of \cite{Gottesman}).
(It is possible to have sets of non-null $X$-vectors that are 
not linearly independent; however, as discussed in \cite{Gottesman}, 
the generators can always be transformed so that they consist of primary 
and secondary types.)

To choose the codewords, we augment these generators with a set of 
$k$ {\em seed\/} generators (where $k = n-d$), which are chosen so that: \nl
(a) the $X$-vectors of the seed and primary generators are linearly 
independent; and \nl
(b) each seed generator commutes with each secondary generator. \nl
Let $M_1,\ldots,M_b$, $L_1,\ldots,L_r$, and $N_1, \ldots, N_k$ 
be the primary, secondary, and seed generators, respectively 
(where $b+r = d$).
Then, as shown in \cite{Gottesman}, each $k$-qubit basis state 
$|c_1 \ldots c_k\ket$ can be associated with a quantum codeword 
\begin{equation}
{\textstyle{ 1 \over \sqrt{2^b} } }
\sum_{a_1 \ldots a_b \in \{0,1\}^b} 
M_1^{a_1} \cdots M_b^{a_b} 
N_1^{c_1} \cdots N_k^{c_k} |\overbrace{0 \ldots 0}^{n}\,\ket,
\label{main}
\end{equation}
(and, by linear extension, this defines the codeword for an arbitrary 
$k$-qubit state).
Note that these $2^k$ basic codewords are all valid quantum states and are 
mutually orthogonal, since condition (a) implies that the states 
$M_1^{a_1} \cdots M_b^{a_b} N_1^{c_1} \cdots N_k^{c_k} 
|0 \ldots 0\ket$ for $a_1,\ldots,a_b,c_1,\ldots,c_k \in \01$ are 
all distinct basis states.
Also, the basic codewords are all fixed by the stabilizer 
(i.e.,\ they lie within the specified code).
To see why this is so, it is useful to note that 
\begin{equation}
{\textstyle{ 1 \over \sqrt{2^b} } }
(I + M_1) \cdots (I + M_b)
N_1^{c_1} \cdots N_k^{c_k} |\overbrace{0 \ldots 0}^n\,\ket
\label{main2}
\end{equation}
is equivalent to (\ref{main}).
Expression (\ref{main2}) is fixed by each primary generator $M_i$, because 
${M_i\cdot(I+M_i)}=(M_i+I)$, and is fixed by each secondary 
generator $L_j$, because 
\begin{eqnarray}
\lefteqn{L_j \cdot 
\left({\textstyle{ 1 \over \sqrt{2^b} } }
(I + M_1) \cdots (I + M_b)
N_1^{c_1} \cdots N_k^{c_k} |0 \ldots 0\ket
\right)} \hspace*{40mm} \nonumber \\
& = & {\textstyle{ 1 \over \sqrt{2^b} } }
(I + M_1) \cdots (I + M_b)
N_1^{c_1} \cdots N_k^{c_k} \cdot L_j \, |0 \ldots 0\ket \nonumber \\
& = & {\textstyle{ 1 \over \sqrt{2^b} } }
(I + M_1) \cdots (I + M_b)
N_1^{c_1} \cdots N_k^{c_k} |0 \ldots 0\ket,
\nonumber
\end{eqnarray}
where we are using the fact that $L_j$ commutes with each seed generator 
(condition (b)).

In \cite{Gottesman}, methods for constructing generators and seeds for 
a variety of codes are given, including the following seeds for the 
eight qubit code (\ref{eight0}), (\ref{eight1}):
\begin{eqnarray}
N_1 & = & X \ox X \ox \I  \ox \I  \ox \I  \ox \I  \ox \I  \ox \I \nonumber\\
N_2 & = & X \ox \I \ox X  \ox \I  \ox \I  \ox \I  \ox \I  \ox \I 
\label{seed8} \\
N_3 & = & X \ox \I \ox \I  \ox \I  \ox X  \ox \I  \ox \I  \ox \I \nonumber.
\end{eqnarray}
It has been shown \cite{Beckman} that any set of primary and secondary 
generators can be efficiently augmented with seed generators.
We shall present a related method for constructing seed generators that 
takes advantage of the generators being in a special form.

The resulting basic codewords for the eight qubit code are written out in 
full in \cite{Gottesman,St2}.
For large codes, it is impractical to explicitly write out the basic 
codewords.  For instance, one of the codes from \cite{Gottesman}
is a sixteen qubit code that protects ten qubits against one error.
The code is described by six generators and ten seed generators.
If written out in full, this code consists of 1,024 codewords, each of 
which is a superposition of 32 basis states (which amounts to 32,768 basis 
states in total!).

An alternative is to produce a gate array that transforms basis states 
into codewords.
This, in addition to specifying the code, indicates how encodings might 
be {\em computed\/} by quantum computers.
The orthogonality of the codewords implies that the mapping of $k$-qubit 
states to $n$-qubit codewords can, in principle, be implemented unitarily 
(technically, the unitary transformation would map $n$-qubit states of 
the form $|c_1 \ldots c_k\ket \otimes |\overbrace{0 \ldots 0}^{d}\,\ket$ 
to $n$-qubit codewords).
This does not imply that the unitary transformation can be implemented 
{\em efficiently} by a gate array (for example, with a polynomial number 
of operations with respect to $n$).
A direct conversion of the $2^k$ basic codewords (each of which is a 
superposition of $2^b$ basis states) into a gate array would generally 
result in an exponential number of operations.

The expression (\ref{main2}) is equivalent to (\ref{main}) and is apparently 
simpler in that it contains no exponentially large sums; however, it is not 
clear how to translate this into a gate array, since operations of the form 
${1 \over \sqrt{2}}(I+M_i)$ are not unitary.

\section{Converting Generators into Standard Form}

Let $G_1,\ldots,G_d$ be generators of the stabilizer of some $n$-qubit code.
We shall show how to systematically convert the matrices $X_{G}$ and $Z_{G}$ 
into a {\em standard form} which is very useful for producing 
codewords.  From this form, a set of primary and secondary generators, 
as well as a suitable set of seed generators, are readily available.
More importantly, we shall show in the next section how to convert a set of 
generators in standard form into a gate array of size $O(n d)$ that 
transforms $k$-qubit states (where $k=n-d$) into codewords.

Our conversion will involve transformations which change a generator $G_i$ 
into $G_i \cdot G_j$, where $j \neq i$.
Since the stabilizer is a group, it is unchanged by such a transformation 
on the generators.
In terms of the $X$- and $Z$-matrices, such an operation adds the 
$j$th column to the $i$th column in both matrices (in modulo 2 arithmetic), 
which is a basic step in Gaussian elimination.
We shall also allow the $n$ qubit positions to be reordered, which 
corresponds to a reordering of the rows in both matrices (another basic step 
in Gaussian elimination).
This reordering is not strictly necessary, but is convenient for 
notational purposes; it clearly does not change the characteristics of 
the code.

Let $X^{(0)}$ and $Z^{(0)}$ denote the original $X$- and $Z$-matrices 
($X_{G}$ and $Z_{G}$) of the generators (they are each $n \times d$ 
matrices).
We shall perform a suitable Gaussian elimination on these matrices, and 
then augment them with columns corresponding to seed generators leading 
to what we shall call a {\em standard form}.

By performing Gaussian elimination on $X^{(0)}$, 
we can obtain matrices of the form 
$$
X^{(1)} = \bordermatrix{ 
    & \overbrace{}^r & \overbrace{}^b \cr
\mbox{\scriptsize $r+k$}\, \big\{ 
& \mbox{\LARGE 0} & \mbox{\LARGE $A$} \cr
\hspace*{6mm} \mbox{\scriptsize $b$}\, \big\{ 
& \mbox{\LARGE 0} & \mbox{\LARGE $I$} \cr
}
\hspace*{30mm}
Z^{(1)} = \bordermatrix{ 
    & \overbrace{}^r & \overbrace{}^b \cr
\mbox{\scriptsize $r+k$}\, \big\{ 
& \mbox{\LARGE $B$} & \mbox{\LARGE $C$\,} \cr
\hspace*{6mm} \mbox{\scriptsize $b$}\, \big\{ 
& \mbox{\LARGE $D$} & \mbox{\LARGE $E$} \cr
},
$$
where $b$ is the rank of $X^{(0)}$, $r = d-b$, and $k = n-d$.
At this stage, the $Z$-matrix $Z^{(1)}$ has no special form.
Next, by performing Gaussian elimination on the first $r+k$ rows and 
the first $r$ columns of $Z^{(1)}$, we can transform $B$, 
the $(r+k) \times r$ submatrix of $Z^{(1)}$, into $B^{^{\prime}}$ of the form 
$$
\mbox{\large $B^{^{\prime}}$} = \bordermatrix{ 
    & \overbrace{}^{r_2} & \overbrace{}^{r_1} \cr
\hspace*{1.2mm} 
\mbox{\scriptsize $k$}\, \big\{ & \mbox{\Large 0} & \mbox{\Large $B_1$} \cr
\mbox{\scriptsize $r_2$}\, \big\{ & \mbox{\Large 0} & \mbox{\Large $B_2$}\cr
\mbox{\scriptsize $r_1$}\, \big\{ & \mbox{\Large 0} & \mbox{\Large $I$} \cr
},
$$
where $r_1$ is the rank of $B$ and $r_2 = r-r_1$.
Note that this does not affect the last $b$ rows or the first $r$ columns 
of $X^{(1)}$.
Thus, the resulting forms of the $X$- and $Z$-matrices 
(blocked with the new partition) are 
$$
X^{(2)} = \bordermatrix{ 
    & \overbrace{}^{r_2} & \overbrace{}^{r_1} & \overbrace{}^b \cr
\hspace*{1.2mm} \mbox{\scriptsize $k$}\, \big\{ 
& \mbox{\Large 0} & \mbox{\Large 0} & \mbox{\it\Large $A_1$} \cr
\mbox{\scriptsize $r_2$}\, \big\{ 
& \mbox{\Large 0} & \mbox{\Large 0} & \mbox{\it\Large $A_2$} \cr
\mbox{\scriptsize $r_1$}\, \big\{ 
& \mbox{\Large 0} & \mbox{\Large 0} & \mbox{\it\Large $A_3$} \cr
\hspace*{1.6mm} \mbox{\scriptsize $b$}\, \big\{ 
& \mbox{\Large 0} & \mbox{\Large 0} & \mbox{\Large $I$} \cr
}
\hspace*{20mm}
Z^{(2)} = \bordermatrix{ 
    & \overbrace{}^{r_2} & \overbrace{}^{r_1} & \overbrace{}^b \cr
\hspace*{1.2mm} \mbox{\scriptsize $k$}\, \{ 
& \mbox{\Large 0} & \mbox{\Large $B_1$} & \mbox{\it\Large $C_1$} \cr
\mbox{\scriptsize $r_2$}\, \{ 
& \mbox{\Large 0} & \mbox{\Large $B_2$} & \mbox{\it\Large $C_2$} \cr
\mbox{\scriptsize $r_1$}\, \{ 
& \mbox{\Large 0} & \mbox{\Large $I$} & \mbox{\it\Large $C_3$} \cr
\hspace*{1.6mm} \mbox{\scriptsize $b$}\, \{ 
& \mbox{\Large $D_1$} & \mbox{\Large $D_2$} & \mbox{\Large $E$} \cr
}.
$$
The first $r = r_1 + r_2$ columns correspond to secondary generators, 
and the last $b$ columns correspond to primary generators (it is clear 
that the last $b$ columns of $X^{(2)}$ are linearly independent).

Next, we shall augment these matrices with $k$ columns corresponding to 
$k$ seed generators.
Recall that the properties that the seed generators must have are: 
their $X$-vectors are linearly independent of those of the primary 
generators; and, they commute with the secondary generators.
Let the $X$- and $Z$-matrices of the seed generators be 
$$
X^{(s)} = \bordermatrix{ 
    & \overbrace{}^{k} \cr
\hspace*{1.2mm} \mbox{\scriptsize $k$}\, \big\{ & \mbox{\Large $I$} \cr
\mbox{\scriptsize $r_2$}\, \big\{ & \mbox{\Large $0$} \cr
\mbox{\scriptsize $r_1$}\, \big\{ & 
\mbox{\Large $B_1^{^{\mbox{\scriptsize $T$}}}$} \cr
\hspace*{1.6mm} \mbox{\scriptsize $b$}\, \big\{ & \mbox{\Large $0$} \cr
}
\hspace*{30mm}
Z^{(s)} = \bordermatrix{ 
    & \overbrace{}^{k} \cr
\hspace*{1.2mm} \mbox{\scriptsize $k$}\, \big\{ & \mbox{\Large $0$} \cr
\mbox{\scriptsize $r_2$}\, \big\{ & \mbox{\Large $0$} \cr
\mbox{\scriptsize $r_1$}\, \big\{ & \mbox{\Large $0$} \cr
\hspace*{1.6mm} \mbox{\scriptsize $b$}\, \big\{ & \mbox{\Large $0$} \cr
},
$$
where $B_1^T$ is the transpose of $B_1$.
It is clear that the $X$-vectors of these seed generators are linearly 
independent of those of the primary generators (that is, the columns of 
$X^{(s)}$ are linearly independent of the last $b$ columns of $X^{(2)}$).
To see why these seed generators commute with the secondary generators, 
note that this condition is equivalent to having an even number of 1's 
in their $X$-vectors in common with the $Z$-vector of each secondary 
generator.
This is equivalent to
$$
\bordermatrix{ 
& \overbrace{}^{k} & \overbrace{}^{r_2} & \overbrace{}^{r_1} 
& \overbrace{}^{b} \cr
\mbox{\scriptsize $k$}\, \{ & \mbox{\Large $I$} & \mbox{\Large $0$}  
& \mbox{\Large $B_1$} & \mbox{\Large $0$} \cr
} 
\bordermatrix{
& \overbrace{}^{r_2} & \overbrace{}^{r_1} \cr
& \mbox{\Large $0$} & \mbox{\Large $B_1$} \cr
& \mbox{\Large $0$} & \mbox{\Large $B_2$} \cr
& \mbox{\Large $0$} & \mbox{\Large $I$} \cr
& \mbox{\Large $0$} & \mbox{\Large $D_2$} \cr
}
\ \ = \ \ 
\bordermatrix{ 
& \overbrace{}^{r_2} & \overbrace{}^{r_1} \cr
\mbox{\scriptsize $k$}\, \{ & \mbox{\Large $0$} & \mbox{\Large $0$} \cr
}, 
$$
which holds because $B_1 + B_1 = 0$ (modulo 2).

The augmented matrices $X^{(*)}$ and $Z^{(*)}$, which include the 
seed, secondary, and primary generators, are as follows.
\begin{equation}
X^{(*)} = \bordermatrix{ 
& \overbrace{}^k & \overbrace{}^{r_2} & \overbrace{}^{r_1} &\overbrace{}^b\cr
\hspace*{1.2mm} \mbox{\scriptsize $k$}\, \big\{ 
& \mbox{\Large $I$} & \mbox{\Large $0$} & \mbox{\Large $0$} 
& \mbox{\it\Large $A_1$} \cr
\mbox{\scriptsize $r_2$}\, \big\{ 
& \mbox{\Large $0$} & \mbox{\Large $0$} & \mbox{\Large $0$} 
& \mbox{\it\Large $A_2$} \cr
\mbox{\scriptsize $r_1$}\, \big\{ 
& \mbox{\Large $B_1^{^{\mbox{\scriptsize $T$}}}$} & \mbox{\Large $0$} 
& \mbox{\Large $0$} & \mbox{\it\Large $A_3$} \cr
\hspace*{1.6mm} \mbox{\scriptsize $b$}\, \big\{ 
& \mbox{\Large $0$} & \mbox{\Large $0$} & \mbox{\Large $0$} 
& \mbox{\Large $I$} \cr
}
\hspace*{15mm}
Z^{(*)} = \bordermatrix{ 
& \overbrace{}^k & \overbrace{}^{r_2} & \overbrace{}^{r_1} &\overbrace{}^b\cr
\hspace*{1.2mm} \mbox{\scriptsize $k$}\, \{ 
& \mbox{\Large $0$} & \mbox{\Large 0} & \mbox{\Large $B_1$} 
& \mbox{\it\Large $C_1$} \cr
\mbox{\scriptsize $r_2$}\, \{ 
& \mbox{\Large $0$} & \mbox{\Large 0} & \mbox{\Large $B_2$} 
& \mbox{\it\Large $C_2$} \cr
\mbox{\scriptsize $r_1$}\, \{ 
& \mbox{\Large $0$} & \mbox{\Large 0} & \mbox{\Large $I$} 
& \mbox{\it\Large $C_3$} \cr
\hspace*{1.6mm} \mbox{\scriptsize $b$}\, \{ 
& \mbox{\Large $0$} & \mbox{\Large $D_1$} & \mbox{\Large $D_2$} 
& \mbox{\Large $E$} \cr
}
\label{standard}
\end{equation}
Call any specification of generators in the above form a 
{\it standard form}.
Note that there are $O(n d)$ 1's in each matrix.

As an example, consider the generators for the eight qubit code described 
in the previous section (represented by the matrices of equations 
(\ref{eight1})).
A standard form for this code is 
\begin{equation}
X^{(*)} = 
\pmatrix{1 & 0 & 0 & 0 & 1 & 1 & 1 & 0 \vspace*{-4mm} \cr
         0 & 1 & 0 & 0 & 1 & 1 & 0 & 1 \vspace*{-4mm} \cr
         0 & 0 & 1 & 0 & 1 & 0 & 1 & 1 \vspace*{-4mm} \cr
         1 & 1 & 1 & 0 & 0 & 1 & 1 & 1 \vspace*{-4mm} \cr
         0 & 0 & 0 & 0 & 1 & 0 & 0 & 0 \vspace*{-4mm} \cr
         0 & 0 & 0 & 0 & 0 & 1 & 0 & 0 \vspace*{-4mm} \cr
         0 & 0 & 0 & 0 & 0 & 0 & 1 & 0 \vspace*{-4mm} \cr
         0 & 0 & 0 & 0 & 0 & 0 & 0 & 1 \cr}
\hspace*{10mm}
Z^{(*)} = 
\pmatrix{0 & 0 & 0 & 1 & 0 & 0 & 0 & 0 \vspace*{-4mm} \cr
         0 & 0 & 0 & 1 & 0 & 1 & 0 & 1 \vspace*{-4mm} \cr
         0 & 0 & 0 & 1 & 1 & 0 & 1 & 0 \vspace*{-4mm} \cr
         0 & 0 & 0 & 1 & 1 & 1 & 0 & 0 \vspace*{-4mm} \cr
         0 & 0 & 0 & 1 & 1 & 1 & 1 & 1 \vspace*{-4mm} \cr
         0 & 0 & 0 & 1 & 1 & 0 & 0 & 1 \vspace*{-4mm} \cr
         0 & 0 & 0 & 1 & 0 & 1 & 1 & 0 \vspace*{-4mm} \cr
         0 & 0 & 0 & 1 & 0 & 0 & 1 & 1 \cr}
\label{eight2}
\end{equation}
(with $k=3$, $r_1=1$, $r_2=0$, and $b=4$), where the fourth and fifth qubit 
positions (rows) have been transposed in the Gaussian elimination process.
Modulo this transposition, the last five columns generate the same 
stabilizer as the original five generators.
The first three columns correspond to seed generators, the fourth column 
corresponds to a secondary generator, and the last four columns correspond 
to primary generators.
The seed generators are not the same as those presented in Eq.~(\ref{seed8}), 
so the basis codewords will be different.  However, since the stabilizer is 
the same, the full coding space is the same as before.

\section{Construction of Gate Arrays}

We shall construct a $n$-qubit gate array with $O(n d)$ operations that 
computes the mapping 
\begin{equation}
|c_1 \ldots c_k\ket \otimes |\overbrace{0 \ldots 0}^d\,\ket 
\mapsto 
{\textstyle{{1 \over \sqrt{2^b}} }} \sum_{a_1 \ldots a_b \in \01^b} 
M_1^{a_1} \cdots M_b^{a_b} 
N_1^{c_1} \cdots N_k^{c_k} |\overbrace{0 \ldots 0}^n\,\ket.
\label{mapping}
\end{equation}
This mapping (\ref{mapping}) is the composition of two mappings, 
\begin{equation}
|c_1 \ldots c_k\ket \otimes |\overbrace{0 \ldots 0}^d\,\ket 
\mapsto 
|c_1 \ldots c_k\ket \ox |\overbrace{0 \ldots 0}^r\,\ket \ox 
{\textstyle{{1 \over \sqrt{2^b}} }} \sum_{a_1 \ldots a_b \in \01^b} 
|a_1 \ldots a_b\ket 
\label{pre}
\end{equation}
and
\begin{equation}
|c_1 \ldots c_k\ket \ox |\overbrace{0 \ldots 0}^r\,\ket 
\ox |a_1 \ldots a_b\ket 
\mapsto 
M_1^{a_1} \cdots M_b^{a_b} 
N_1^{c_1} \cdots N_k^{c_k} |\overbrace{0 \ldots 0}^n\,\ket.
\label{post}
\end{equation}

The first mapping (\ref{pre}) is trivially computed by independently 
applying a $Q$ operation to each of the last $b$ qubits, 
where 
$$Q = {\textstyle{1 \over \sqrt{2}}}
\pmatrix{1 & \ 1 \cr 1 & -1}.$$

To compute the second mapping (\ref{post}), it is helpful to consider the 
eight qubit code, whose standard form is given by (\ref{eight2}) in the 
previous section.
Recall that the first three columns specify the seed generators and 
the last four columns specify the primary generators.
Let us temporarily consider a simplified version of the generators 
specified in (\ref{eight2}), where phase shifts are ignored.
This is equivalent to keeping $X^{(*)}$ as is and setting $Z^{(*)}$ 
to all zeros (thus, $Z$'s become $I$'s, $Y$'s become $X$'s, 
and $X$'s and $I$'s are unchanged in the tensor products 
that make up the generators).
With respect to these generators, the mapping (\ref{post}) is given 
by matrix $X^{(*)}$ as 
\begin{equation}
\pmatrix{c_1 \vspace*{-4mm} \cr
         c_2 \vspace*{-4mm} \cr
         c_3 \vspace*{-4mm} \cr
         0   \vspace*{-4mm} \cr
         a_1 \vspace*{-4mm} \cr
         a_2 \vspace*{-4mm} \cr
         a_3 \vspace*{-4mm} \cr
         a_4 \cr}
\mapsto
\pmatrix{1 & 0 & 0 & 0 & 1 & 1 & 1 & 0 \vspace*{-4mm} \cr
         0 & 1 & 0 & 0 & 1 & 1 & 0 & 1 \vspace*{-4mm} \cr
         0 & 0 & 1 & 0 & 1 & 0 & 1 & 1 \vspace*{-4mm} \cr
         1 & 1 & 1 & 0 & 0 & 1 & 1 & 1 \vspace*{-4mm} \cr
         0 & 0 & 0 & 0 & 1 & 0 & 0 & 0 \vspace*{-4mm} \cr
         0 & 0 & 0 & 0 & 0 & 1 & 0 & 0 \vspace*{-4mm} \cr
         0 & 0 & 0 & 0 & 0 & 0 & 1 & 0 \vspace*{-4mm} \cr
         0 & 0 & 0 & 0 & 0 & 0 & 0 & 1 \cr}
\pmatrix{c_1 \vspace*{-4mm} \cr
         c_2 \vspace*{-4mm} \cr
         c_3 \vspace*{-4mm} \cr
         0   \vspace*{-4mm} \cr
         a_1 \vspace*{-4mm} \cr
         a_2 \vspace*{-4mm} \cr
         a_3 \vspace*{-4mm} \cr
         a_4 \cr},
\label{eight3}
\end{equation}
where we are denoting the input state 
$|c_1 c_2 c_3\ket \ox |0\ket \ox |a_1 a_2 a_3 a_4\ket$ 
and the output state as column vectors (the phases are all 
$+1$ here).
This mapping is computed by the following gate array (where we are using 
notation of \cite{G9}).

\setlength{\unitlength}{0.08cm}
\hspace*{65mm}
\begin{picture}(90,90)(20,0)

\put(0,10){\line(1,0){80}}
\put(0,20){\line(1,0){80}}
\put(0,30){\line(1,0){80}}
\put(0,40){\line(1,0){80}}
\put(0,50){\line(1,0){80}}
\put(0,60){\line(1,0){80}}
\put(0,70){\line(1,0){80}}
\put(0,80){\line(1,0){80}}

\put(10,48){\line(0,1){32}}
\put(10,50){\circle{4}}
\put(10,80){\circle*{2}}

\put(20,48){\line(0,1){22}}
\put(20,50){\circle{4}}
\put(20,70){\circle*{2}}

\put(30,48){\line(0,1){12}}
\put(30,50){\circle{4}}
\put(30,60){\circle*{2}}

\put(40,40){\line(0,1){42}}
\put(40,40){\circle*{2}}
\put(40,60){\circle{4}}
\put(40,70){\circle{4}}
\put(40,80){\circle{4}}

\put(50,30){\line(0,1){52}}
\put(50,30){\circle*{2}}
\put(50,50){\circle{4}}
\put(50,70){\circle{4}}
\put(50,80){\circle{4}}

\put(60,20){\line(0,1){62}}
\put(60,20){\circle*{2}}
\put(60,50){\circle{4}}
\put(60,60){\circle{4}}
\put(60,80){\circle{4}}

\put(70,10){\line(0,1){62}}
\put(70,10){\circle*{2}}
\put(70,50){\circle{4}}
\put(70,60){\circle{4}}
\put(70,70){\circle{4}}

\put(-7,9){$a_4$}
\put(-7,19){$a_3$}
\put(-7,29){$a_2$}
\put(-7,39){$a_1$}
\put(-5,48.5){$0$}
\put(-7,59){$c_3$}
\put(-7,69){$c_2$}
\put(-7,79){$c_1$}

\put(20,0){\small Figure 1}

\end{picture}

\ni By inspection, we can see how the operations in the gate array 
correspond to the entries of matrix $X^{(*)}$.
What we have done is to apply each column conditioned on one of its elements.
For instance, if $a_3 = 1$, we apply the third primary generator
by flipping the first, third, and fourth qubits.  We can only do this if
the seventh qubit, which initially has the value $a_3$, has not been
changed before we get around to applying that generator.  If an earlier
column had a 1 in the seventh row, that qubit might have been changed
before it could be used.  This is why we
needed to put the X-matrix in standard form --- each column of the matrix
must be conditioned on the input of the corresponding qubit, so the
diagonal elements must be first in their rows.  A matrix in standard
form has this property (at least for columns not corresponding to secondary
generators), while a more general matrix does not.

Returning to our goal of implementing the mapping (\ref{post}) 
{\em with phase shifts included}, we make the following observations 
about phase shifts and conditional phase shifts acting on basis states 
with phase $\pm 1$. \nl
(a) The action of a conditional phase shift or conditional negation 
depends on the state, but \nl
\hspace*{5.5mm} not on the phase of the state, to which it 
is applied. \nl
(b) The action of a phase shift, or conditional phase shift, may affect 
the phase, but does \nl
\hspace*{5.5mm} not change the state on which it acts. \nl
This implies that the phase shifts can be essentially inserted into 
Figure 1 as they occur in matrix $Z^{(*)}$; they do not affect the evolution 
of the state, but they do affect an evolving phase factor of $\pm 1$.
In the case of the eight qubit code, we insert conditional phase shifts 
into Figure 1 in accordance with the matrix $Z^{(*)}$ of (\ref{eight2}) 
to obtain the following gate array.

\setlength{\unitlength}{0.08cm}
\hspace*{65mm}
\begin{picture}(100,90)(20,0)

\put(-10,10){\line(1,0){7}}
\put(3,10){\line(1,0){54}}
\put(63,10){\line(1,0){17}}
\put(-3,7){\framebox(6,6){$Z$}}
\put(57,7){\framebox(6,6){$Z$}}

\put(-10,20){\line(1,0){7}}
\put(3,20){\line(1,0){44}}
\put(53,20){\line(1,0){27}}
\put(-3,17){\framebox(6,6){$Z$}}
\put(47,17){\framebox(6,6){$Z$}}

\put(-10,30){\line(1,0){47}}
\put(43,30){\line(1,0){24}}
\put(73,30){\line(1,0){7}}
\put(37,27){\framebox(6,6){$Z$}}
\put(67,27){\framebox(6,6){$Z$}}

\put(-10,40){\line(1,0){7}}
\put(3,40){\line(1,0){44}}
\put(53,40){\line(1,0){4}}
\put(63,40){\line(1,0){4}}
\put(73,40){\line(1,0){7}}
\put(-3,37){\framebox(6,6){$Z$}}
\put(47,37){\framebox(6,6){$Z$}}
\put(57,37){\framebox(6,6){$Z$}}
\put(67,37){\framebox(6,6){$Z$}}

\put(-10,50){\line(1,0){47}}
\put(43,50){\line(1,0){4}}
\put(53,50){\line(1,0){27}}
\put(37,47){\framebox(6,6){$Z$}}
\put(47,47){\framebox(6,6){$Y$}}

\put(-10,60){\line(1,0){47}}
\put(43,60){\line(1,0){14}}
\put(63,60){\line(1,0){17}}
\put(37,57){\framebox(6,6){$Y$}}
\put(57,57){\framebox(6,6){$Y$}}

\put(-10,70){\line(1,0){57}}
\put(53,70){\line(1,0){14}}
\put(73,70){\line(1,0){7}}
\put(47,67){\framebox(6,6){$Y$}}
\put(67,67){\framebox(6,6){$Y$}}

\put(-10,80){\line(1,0){90}}

\put(10,48){\line(0,1){32}}
\put(10,50){\circle{4}}
\put(10,80){\circle*{2}}

\put(20,48){\line(0,1){22}}
\put(20,50){\circle{4}}
\put(20,70){\circle*{2}}

\put(30,48){\line(0,1){12}}
\put(30,50){\circle{4}}
\put(30,60){\circle*{2}}

\put(40,33){\line(0,1){14}}
\put(40,53){\line(0,1){4}}
\put(40,63){\line(0,1){19}}
\put(40,40){\circle*{2}}
\put(40,70){\circle{4}}
\put(40,80){\circle{4}}

\put(50,23){\line(0,1){14}}
\put(50,43){\line(0,1){4}}
\put(50,53){\line(0,1){14}}
\put(50,73){\line(0,1){9}}
\put(50,30){\circle*{2}}
\put(50,80){\circle{4}}

\put(60,13){\line(0,1){24}}
\put(60,43){\line(0,1){14}}
\put(60,63){\line(0,1){19}}
\put(60,20){\circle*{2}}
\put(60,50){\circle{4}}
\put(60,80){\circle{4}}

\put(70,10){\line(0,1){17}}
\put(70,33){\line(0,1){4}}
\put(70,43){\line(0,1){24}}
\put(70,10){\circle*{2}}
\put(70,50){\circle{4}}
\put(70,60){\circle{4}}

\put(-17,9){$a_4$}
\put(-17,19){$a_3$}
\put(-17,29){$a_2$}
\put(-17,39){$a_1$}
\put(-15,48.5){$0$}
\put(-17,59){$c_3$}
\put(-17,69){$c_2$}
\put(-17,79){$c_1$}

\put(20,0){\small Figure 2}

\end{picture}

\ni Of course, the fact that $X \cdot Z = Y$ is used here.
We have done essentially the same thing here as with the X-matrix, 
conditioning the phases on specific qubits.
Notice, for example, that the phase factors conditioned on $a_1$ 
correspond to the generator of the fifth column of $Z^{(*)}$.
Also, recall that we exclude the secondary generators, so that the fourth 
column of $Z^{(*)}$ is ignored.
Strictly speaking, the initial $Z$'s for $a_1$, $a_3$, and $a_4$ are also
conditioned on those qubits.  However, since $Z$ has no effect on 
$|0\rangle$, we can apply it as an unconditional operation.

The gate array of Figure 2 computes the mapping (\ref{post}) for the eight 
qubit code, and such a gate array can be constructed in general from 
any matrices $X^{(*)}$ and $Z^{(*)}$ in standard form.

Finally, a full gate array for computing codewords is obtained by 
composing the operations for mappings (\ref{pre}) and (\ref{post}).
In the case of the eight qubit code, we obtain the following (where 
$R = Q \cdot Z$).

\setlength{\unitlength}{0.08cm}
\hspace*{60mm}
\begin{picture}(130,90)(20,0)

\put(-10,10){\line(1,0){7}}
\put(3,10){\line(1,0){54}}
\put(63,10){\line(1,0){17}}
\put(-3,7){\framebox(6,6){$R$}}
\put(57,7){\framebox(6,6){$Z$}}

\put(-10,20){\line(1,0){7}}
\put(3,20){\line(1,0){44}}
\put(53,20){\line(1,0){27}}
\put(-3,17){\framebox(6,6){$R$}}
\put(47,17){\framebox(6,6){$Z$}}

\put(-10,30){\line(1,0){7}}
\put(3,30){\line(1,0){34}}
\put(43,30){\line(1,0){24}}
\put(73,30){\line(1,0){7}}
\put(-3,27){\framebox(6,6){$Q$}}
\put(37,27){\framebox(6,6){$Z$}}
\put(67,27){\framebox(6,6){$Z$}}

\put(-10,40){\line(1,0){7}}
\put(3,40){\line(1,0){44}}
\put(53,40){\line(1,0){4}}
\put(63,40){\line(1,0){4}}
\put(73,40){\line(1,0){7}}
\put(-3,37){\framebox(6,6){$R$}}
\put(47,37){\framebox(6,6){$Z$}}
\put(57,37){\framebox(6,6){$Z$}}
\put(67,37){\framebox(6,6){$Z$}}

\put(-10,50){\line(1,0){47}}
\put(43,50){\line(1,0){4}}
\put(53,50){\line(1,0){27}}
\put(37,47){\framebox(6,6){$Z$}}
\put(47,47){\framebox(6,6){$Y$}}

\put(-10,60){\line(1,0){47}}
\put(43,60){\line(1,0){14}}
\put(63,60){\line(1,0){17}}
\put(37,57){\framebox(6,6){$Y$}}
\put(57,57){\framebox(6,6){$Y$}}

\put(-10,70){\line(1,0){57}}
\put(53,70){\line(1,0){14}}
\put(73,70){\line(1,0){7}}
\put(47,67){\framebox(6,6){$Y$}}
\put(67,67){\framebox(6,6){$Y$}}

\put(-10,80){\line(1,0){90}}

\put(10,48){\line(0,1){32}}
\put(10,50){\circle{4}}
\put(10,80){\circle*{2}}

\put(20,48){\line(0,1){22}}
\put(20,50){\circle{4}}
\put(20,70){\circle*{2}}

\put(30,48){\line(0,1){12}}
\put(30,50){\circle{4}}
\put(30,60){\circle*{2}}

\put(40,33){\line(0,1){14}}
\put(40,53){\line(0,1){4}}
\put(40,63){\line(0,1){19}}
\put(40,40){\circle*{2}}
\put(40,70){\circle{4}}
\put(40,80){\circle{4}}

\put(50,23){\line(0,1){14}}
\put(50,43){\line(0,1){4}}
\put(50,53){\line(0,1){14}}
\put(50,73){\line(0,1){9}}
\put(50,30){\circle*{2}}
\put(50,80){\circle{4}}

\put(60,13){\line(0,1){24}}
\put(60,43){\line(0,1){14}}
\put(60,63){\line(0,1){19}}
\put(60,20){\circle*{2}}
\put(60,50){\circle{4}}
\put(60,80){\circle{4}}

\put(70,10){\line(0,1){17}}
\put(70,33){\line(0,1){4}}
\put(70,43){\line(0,1){24}}
\put(70,10){\circle*{2}}
\put(70,50){\circle{4}}
\put(70,60){\circle{4}}

\put(-15,48.5){$0$}
\put(-15,38.5){$0$}
\put(-15,28.5){$0$}
\put(-15,18.5){$0$}
\put(-15,8.5){$0$}

\put(-28.5,68.7){input $\left\{\matrix{ \vspace*{13mm} \cr}\right.$}

\put(75.5,43.6){
$\left.\matrix{ \ \vspace*{53mm} \cr} \right\}$ codeword}

\put(20,0){\small Figure 3}

\end{picture}

\ni In general, the total number of two-qubit operations is bounded by 
$$r_1 k + (n-1) b \le (n-1) d$$
and the number of one-qubit operations is bounded by $b \le d$.
Thus, the total number of operations is bounded by $n d$, which is 
essentially the length of the description of the generators of the 
stabilizer.
For the eight qubit code, there are four one-qubit operations and 23
two-qubit operations.
For the sixteen qubit code mentioned in Section II, the gate array 
will have no more than 96 operations.

To recover messages from their codewords, it suffices to apply the 
codeword error-correction scheme proposed in \cite{DiSh} 
(without fault-tolerance, if not required) followed by the gate array for 
encoding (e.g.\ Figure 3) backwards.

\acknowledgements

We would like to thank David Beckman, Sam Braunstein, David DiVincenzo, and 
Andrew Steane for helpful discussions about the computation of encodings.
We are also grateful to the Institute for Scientific Interchange in Torino, 
and its director, Professor Mario Rasetti, for making possible the 1996 
workshop on Quantum Computation at which some of this work was performed.
R.C. is supported in part by NSERC of Canada.  D.G. is supported by the
U.S. Department of Energy under Grant No. DE-FG03-92-ER40701 and by DARPA
through a grant to ARO.

\end{document}